\documentclass[a4paper,11pt]{article} 
\pdfoutput=1 

\usepackage{jcappub} 
\usepackage[utf8]{inputenc} 

\usepackage{graphicx} 
\usepackage{epstopdf}
\usepackage{subcaption} 
\usepackage{amsmath}
\usepackage{amssymb}
\usepackage{amsfonts}

\makeatletter
\gdef\@fpheader{}
\g@addto@macro\bfseries{\boldmath}
\makeatother

\newcommand\M{M_{Pl}}
\newcommand\R{\mathbb{R}}

\title{\boldmath Anatomy of geometrical destabilization of inflation}

\author{Tomasz Krajewski,}
\author{Krzysztof Turzy\'nski}

\affiliation{Institute of Theoretical Physics, Faculty of Physics, University of Warsaw, \\ Pasteura 5, 02-093 Warsaw, Poland}

\emailAdd{tomasz.krajewski@fuw.edu.pl}
\emailAdd{krzysztof.turzynski@fuw.edu.pl}

\abstract{
We study geometrical destabilization of inflation
with the aim of determining the fate of excited unstable modes.
We use numerical lattice simulations to track
the dynamics of both the inflaton and the spectator field. 
We find that geometrical destabilization
is a short-lived phenomenon
and that a negative feedback loop prevents field
fluctuations from growing indefinitely. As a result,
fields undergoing geometrical destabilization
are merely shifted to a new classical configuration corresponding
to a uniform value of the spectator field within a Hubble patch.
}


\arxivnumber{2205.13487}

\begin{document}
\maketitle
\flushbottom

\section{Introduction}

Cosmological inflation~\cite{Starobinsky:1980te, Sato:1980yn, Guth:1980zm, Linde:1981mu, Albrecht:1982wi, Linde:1983gd} 
has by now become a natural ingredient of the standard cosmological model 
(see e.g.\ \cite{VM} for a pedagogical introduction).
It describes a phase of the evolution of the Universe in which its expansion accelerates and
vacuum quantum fluctuations of the gravitational and matter fields 
are amplified to cosmological perturbations~\cite{Starobinsky:1979ty, Mukhanov:1981xt, Hawking:1982cz,  Starobinsky:1982ee, Guth:1982ec, Bardeen:1983qw},
which later seed the cosmic microwave background (CMB) anisotropies and the large-scale structure of our Universe. 

As inflation is rather a broad theoretical framework than a well-defined theory, there are many open questions about specific points of history
of the inflationary Universe. One of them is related to {\em reheating}, i.e.\ the passage from the inflationary era to radiation-dominated era,
because there is no definitive prescription for coupling the inflationary sector to the Standard Model of particle physics or its extension thereof. 

Nonetheless, already the pioneering works of Starobinsky describe the model of inflation together with reheating
by gravitational particle creation in the regime of the weak narrow parametric resonance \cite{AStar1,AStar2} (see, {\it e.g.}, \cite{DeFelice:rev} for a review).
At present, the observations can be effectively described with a minimal setup: a single scalar inflaton field with canonical kinetic term, minimally coupled to gravity, and evolving in a sufficiently flat 
potential~\cite{Planck18-1,Planck18-2,Martin:2013tda, Martin:2013nzq,Renaux-Petel:2015bja}. 
Inflation is then typically terminated when the inflaton leaves the slow-roll regime. The subsequent advent of the radiation domination era is not related do any distinct observational signatures,
which has led many authors to include the ignorance about that era into theoretical uncertainties in the predictions of inflationary models.

However, many inflationary models come naturally equipped with the possibility 
that an effective classical force associated with the inflaton can lead to non-adiabatic excitations of scalar field fluctuations through parametric resonance,
thereby modifying the evolution of the Universe. 
This view was first advocated 
(for narrow parametric resonance)
in~\cite{BT}, but the effect was later shown negligible \cite{KLS}, 
whereas the amplification of those fluctuations by a broad parametric resonance \cite{KLS,PRH2,PRH3} is still a viable
candidate for reheating. 

This phenomenon of {\em preheating} is typically considered after the slow-roll phase (see, {\it e.g.}, \cite{IR:rev,FE} for a review), but it can also affect the dynamics of the Universe during inflation in the multi-field setup \cite{Battefeld:2011yj}, including a premature
end of inflation. There are also other mechanisms which can end inflation prematurely or change its course. 
In hybrid inflation~\cite{Linde:1991km, Linde:1993cn, Copeland:1994vg, Lyth:1998xn},
there is a `spectator' field with an inflaton-dependent mass. This field develops a tachyonic instability which either quickly terminates inflation or the field takes over the role of the original inflaton field
\cite{Clesse:2010iz, Avgoustidis:2011em, Martin:2011ib}.

In the context of multi-field inflationary models with non-canonical kinetic terms, a possibility of {\em geometrical destabilization} has been proposed \cite{RT}. 
With a negative field-space curvature, an inflationary trajectory can be destabilized by a `geometrical' force dominating over forces originating from the potential.
It has been hinted that geometrical destabilization may either end inflation prematurely \cite{GD1} or trigger a new phase of inflation \cite{Cicoli:2018ccr,GD2}, especially that geometrical
destabilization is a self-constraining phenomenon \cite{Grocholski:2019mot}. 

The goal of our analysis is to provide a definitive answer
to this dilemma by means of state-of-the-art numerical lattice simulations
of the dynamics of the scalar fields. 

Our work is organized as follows. In Section~\ref{sec:s2}, we briefly
describe the basics of geometrical destabilization.
In Section~\ref{sec:results} we introduce specific, representative 
models of
geometrical destabilization and present results of lattice
simulations of scalar field dynamics in these models.
The discussion of our numerical results is presented in Section~\ref{sec:discussion}. Technical remarks about our numerical
lattice simulations are relegated to Appendix~A.
Throughout the paper we adopt natural units with Planck mass $M_P=1$, unless indicated otherwise. 

\section{Rudiments of geometrical destabilization}
\label{sec:s2}

The idea of geometrical destabilization
is based on the possibility that the action for scalar fields
minimally coupled to gravity
contains kinetic terms that can be written as:
\begin{equation}\label{eq:action}
{\cal L}_{\rm kin}=-\frac12 G_{IJ}(\phi^K) g^{\mu \nu} \partial_{\mu} \phi^I \partial_{\nu} \phi^J \,, 
\end{equation}
where the manifold described by the field-space metric $G_{IJ}$ has nonzero curvature.

The dynamics of models of inflation described by the Lagrangian \eqref{eq:action} have been extensively studied in the past two decades 
(see, {\em e.g.}, \cite{Sasaki:1995aw,Mukhanov:1997fw,GrootNibbelink:2001qt}). On a spatially flat Friedmann-Lema\^itre-Robertson-Walker universe, with metric 
\begin{equation}
\mathrm{d}s^2=-\mathrm{d}t^2+a^2(t)\mathrm{d}{\vec x}^2\,,
\end{equation}
where $t$ is cosmic time and $a(t)$ denotes the scale factor, and with homogeneous scalar fields $ \phi^I$, the equations of motion take the form: 
\begin{eqnarray}
 3H^2 \M^2&=&\frac12 \dot \sigma^2+V\,,  \\
 \dot{H} M_P^2&=&-\frac12\dot \sigma^2\,, \label{Hdot} \\
 \label{eq:scfi_eom}
{\cal D}_t \dot \phi^I  +3H  \dot \phi^I+G^{IJ} V_{,J}&=&0\, .
\end{eqnarray}
In these expressions, dots denote derivatives with respect to $t$, $H \equiv \dot a/a$ is the Hubble parameter, $\tfrac{1}{2}\dot \sigma^2 \equiv \tfrac{1}{2}G_{IJ} \dot \phi^I \dot \phi^J$ is the kinetic energy of the fields, and, hereafter, ${\cal D}_t A^I \equiv \dot{A^I} + \Gamma^I_{JK} \dot \phi^J A^K$ for a field-space vector $A^I$ (field-space indices are lowered and raised with the field-space metric and its inverse respectively).

The behavior of linear fluctuations about such a background is described by the second-order action
 \begin{eqnarray}
S_{(2)}= \int  \mathrm{d}t\, \mathrm{d}^3x \,a^3\left(G_{IJ}\mathcal{D}_tQ^I\mathcal{D}_tQ^J-\frac{1}{a^2}G_{IJ}\partial_i Q^I \partial^i Q^J-M_{IJ}Q^IQ^J\right),
\label{S2}
\end{eqnarray}
where $Q^I$'s are the Mukhanov-Sasaki variables and $M_{IJ}$ is a mass (squared) matrix. The equations of motion following from~\eqref{S2} read:
\begin{eqnarray}
{\cal D}_t {\cal D}_t Q^I  +3H {\cal D}_t Q^I +\frac{k^2}{a^2} Q^I +M^I_{\,J} Q^J=0\,
\label{pert}
\end{eqnarray}
with
\begin{eqnarray} 
\label{masssquared}
M^{I}_{\,J} &=& V^{I}_{; J} - {\mathcal{R}^{I}_{\,KLJ}\dot \phi^K \dot \phi^L} -\frac{1}{a^3 \M^2}\mathcal{D}_t\left(\frac{a^3}{H} \dot \phi^
I \dot \phi_J\right).
\end{eqnarray}
In a curved field space, {\em i.e.}~for non-vanishing $\mathcal{R}^{I}_{\,KLJ}$, the second term in~(\ref{masssquared}) can be negative. As a result, at least some of the field fluctuactions can become
tachyonic, making the inflationary trajectory unstable.
More specifically,
we can rewrite equations of motion \eqref{pert} in the adiabatic-entropic basis $(e_{\sigma}^I,e_{s}^I)$ \cite{Gordon:2000hv,GrootNibbelink:2001qt}, where $e_{\sigma}^I \equiv \dot \phi^I /{\dot \sigma}$ is the unit vector tangent to the background trajectory in field space, and $e_s^I$ is orthonormal to $e_{\sigma}^I$.
The adiabatic perturbation $Q_{\sigma} \equiv e_{\sigma I} Q^I$ is  proportional to the comoving curvature perturbation ${\cal R}=\frac{H}{{\dot \sigma}} Q_{\sigma}$, while
 the entropic fluctuation $Q_s$, perpendicular to the background trajectory, exhibits genuine multi-field effects.

The equation of motion for superhorizon modes of the entropic fluctuation simplifies as well to
\begin{equation}
\ddot Q_s+3 H \dot Q_s +m^{2}_{s {\rm (eff)}} Q_s = 0 \label{eqQs}\,,
\end{equation}
where we denote by $m^{2}_{s {\rm (eff)}}$ the effective entropic mass on superhorizon scales 
\begin{equation}
\frac{m^{2}_{s {\rm (eff)}}}{H^2} \equiv \frac{V_{;ss}}{H^2} +3 \eta_\perp^2+ \epsilon_1  \, \R \M^2 \,.
\label{ms2}
\end{equation}
It contains three contributions: the first one is the usual Hessian of the potential, the second reflects bending of the inflationary trajectory and the third, proportional to the field-space Ricci scalar curvature $\R$, encodes field-space geometrical effects \cite{Peterson:2011yt,Renaux-Petel:2014htw}. 

From \eqref{eqQs} and~\eqref{ms2}, 
the mechanism of geometrical destabilization of inflation is readily identified: it corresponds to situations in which the geometrical contribution is negative and dominates the sum of the two other contributions, so that the entropic fluctuation is tachyonic, and the underlying background trajectory is unstable. As $\epsilon_1$
is a positive quantity, the geometrical destabilization in two-field models can only arise in setups with a scalar curvature that is negative, which is related to the fact that this makes neighbouring geodesics diverge from one another. When the field-space curvature is positive, it renders the entropic fluctuations even more massive, and does not modify the standard picture. Hence, from now in, we shall consider only negatively curved field spaces.

\section{Results}
\label{sec:results}

\subsection{A minimal realization of geometrical destabilization}
\label{sec:minimal}

For our numerical simulations, we use a model of slow-roll inflation driven by a scalar field $\phi$ with canonical kinetic term and potential $V(\phi)$, with the Lagrangian ${\cal L}_{\phi}=-\frac12 (\partial \phi)^2-V(\phi)$, where  
\begin{equation}
    V(\phi) = \Lambda^4 \left( 1 - e^{- \sqrt{\frac{2}{3}} \frac{\phi}{\M}} \right)^2
\end{equation}
is the Starobinsky potential normalized to reproduce the central value of the observed normalization of the curvature perturbations.
In addition, we consider an extra scalar field $\chi$, stabilized at the bottom of its potential by a large mass, larger than the Hubble scale. This is described by the simple Lagrangian ${\cal L}_{\chi}=-\frac12 (\partial \chi)^2-\frac12 m^2 \chi^2$, where $m$ stands for the heavy mass, \textit{i.e.} $m^2 \gg H^2$. 

The dimension six operator describing interactions with the two sector outlined above is ${\cal L}_{{\rm int}}\propto-(\partial \phi)^2 \chi^2/M^2$, where $M$ is a scale of new physics that lies well above the Hubble scale, $M \gg H$. Such an operator respects the (approximate) shift-symmetry of the inflaton and is therefore expected from an effective field theory point of view. Our total Lagrangian thus reads
\begin{equation}
{\cal L}=-\frac12 (\partial \phi)^2  \left(1+2 \frac{\chi^2}{M^2} \right) -V(\phi)-\frac12 (\partial \chi)^2-\frac12 m^2 \chi^2\,.
\label{eq:minimal}
\end{equation}
The dimension six operator generates a curved field space with metric $(1+2 \chi^2/M^2) ({\rm d} \phi)^2+ ({\rm d} \chi)^2$, whose Ricci scalar is negative and reads 
$\R =-4/M^2(1+2 \chi^2/M^2)^{2}$.
Along the inflationary valley $\chi=0$, the entropic fluctuation $Q_s$, which then simply coincides with the fluctuation of $\chi$, thus acquires the effective mass \eqref{ms2}, \textit{i.e.}
\begin{equation}
\label{eq:ms2_2}
m^{2}_{s {\rm (eff)}}=m^2-4 \epsilon_1 H^2 \left( \M/M\right)^2\,,
\end{equation}
as we have here $V_{;ss}=m^2, \left(\R\right)|_{\chi=0}=-4/M^2$, and the inflationary trajectory along $\chi=0$ is a field-space geodesic, so that $\eta_\perp=0$. 

As $\epsilon_1 H^2$ grows during inflation, at the critical point such that 
\begin{equation}
\label{eq:epsilonc}
\epsilon_{1, {\rm c}} =  \frac{1}{4} \left(\frac{m}{H_{\rm c}}\right)^2 \left(\frac{M}{\M}\right)^2\, .
\end{equation}
the effective mass (\ref{eq:ms2_2}) becomes negative, which triggers geometrical destabilization of $\chi$. 
The model has thus two tunable dimensionless parameters, $m^2/H_{\rm c}^2$ and $M^2/\M^2$ whose values control the time at which the geometrical destabilization occurs and the strength of the phenomenon. We shall consider four sets of parameters, listed in Table \ref{tab:pars}. For further convenience, the layout of Table \ref{tab:pars} directly corresponds to the layout of subsequent Figures.

\begin{table}[h]
    \centering
    \begin{tabular}{||ccc|ccc||}
    \hline
    \hline
    $\frac{m_h^2}{H_{\rm c}^2}$ & $\frac{M^2}{\M^2}$ & $\frac{k_{\rm cutoff}}{\M}$ & $\frac{m_h^2}{H_{\rm c}^2}$ & $\frac{M^2}{\M^2}$ & $\frac{k_{\rm cutoff}}{\M}$ \\
    \hline
    \hline
    \multicolumn{3}{||c}{Model 1A} & \multicolumn{3}{|c||}{Model 1B} \\
    $10^2$ & $10^{-3}$ & $320$ & $10^2$ & $10^{-4}$ & $320$ \\
    \hline
    \multicolumn{3}{||c}{Model 2A} & \multicolumn{3}{|c||}{Model 2B} \\
    $10^4$ & $10^{-5}$ & $40$ & $10^4$ & $10^{-6}$ & $40$ \\
    \hline
    \hline
    \multicolumn{3}{||c}{$\epsilon_{1,{\rm c}}=2.5\cdot 10^{-2}$} & \multicolumn{3}{|c||}{$\epsilon_{1,{\rm c}}=2.5\cdot 10^{-3}$} \\
        \multicolumn{3}{||c}{$\Delta N_{\rm c}=4.0$} & \multicolumn{3}{|c||}{$\Delta N_{\rm c}=15.4$} \\
        \hline
        \hline
    \end{tabular}
    \caption{Parameter sets used in the lattice simulations. Models denoted by different numbers correspond to different mass parameters of $\chi$. Models denoted by different letters correspond to different values of $\epsilon_{1,{\rm c}}$, i.e., different onsets of geometrical destabilization, and to different values of $\Delta N_{\rm c}$, i.e., different numbers of e-folds between the onset of geometrical destabilization and the end of single-field inflation.
    Values of different cutoffs $k_{\rm cutoff}$, defined in~Section~\ref{sec:pres} are also shown.}
    \label{tab:pars}
\end{table}

\subsection{Presentation of the lattice simulations results}
\label{sec:pres}

\begin{figure*}[!ht]
\centering
	\begin{subfigure}[t]{0.5\textwidth}
		\label{fig:plot_dump_chi_eta=0.000891}
		\flushleft
		\includegraphics[width=\textwidth]{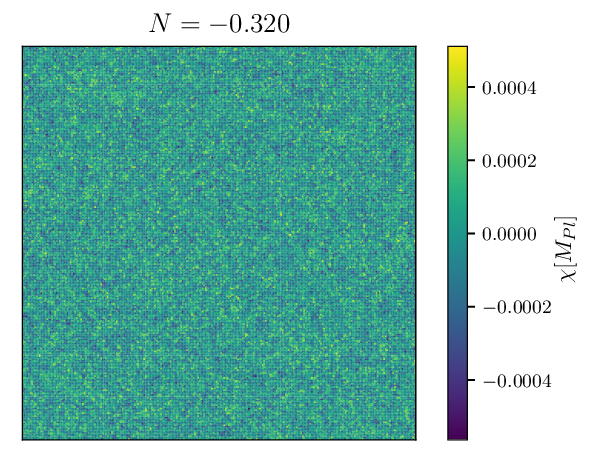}
	\end{subfigure}\hfill %
	\begin{subfigure}[t]{0.5\textwidth}
		\label{fig:plot_dump_chi_eta=47.156771}
		\flushright
		\includegraphics[width=\textwidth]{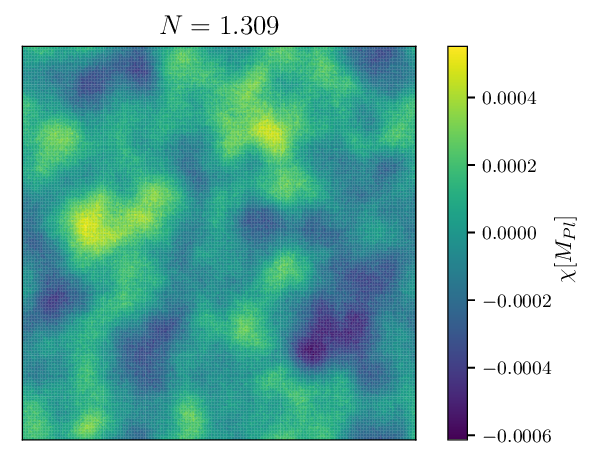}
	\end{subfigure}
    \\
	\begin{subfigure}[t]{0.5\textwidth}
		\label{fig:plot_dump_chi_eta=51.758021}
		\flushleft
		\includegraphics[width=\textwidth]{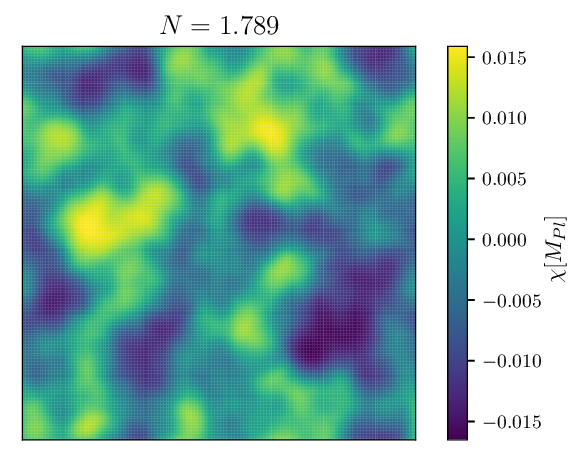}
	\end{subfigure}\hfill %
	\begin{subfigure}[t]{0.5\textwidth}
		\label{fig:plot_dump_chi_eta=52.453381}
		\flushright
		\includegraphics[width=\textwidth]{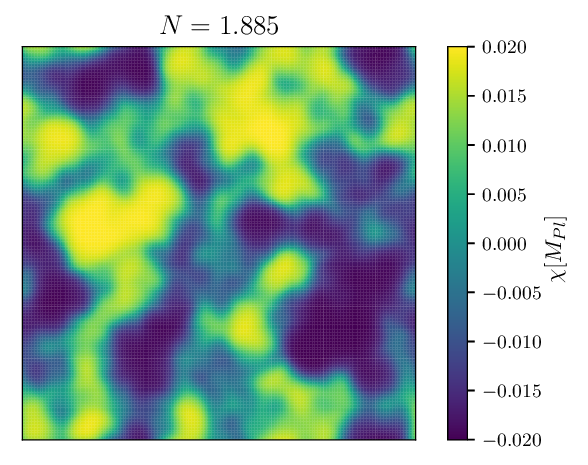}
	\end{subfigure}
	\\
	\begin{subfigure}[t]{0.5\textwidth}
		\label{fig:plot_dump_chi_eta=55.114261}
		\flushleft
		\includegraphics[width=\textwidth]{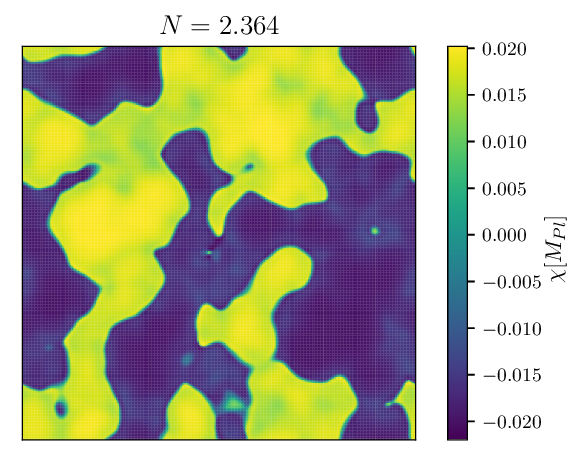}
	\end{subfigure}\hfill %
	\begin{subfigure}[t]{0.5\textwidth}
		\label{fig:plot_dump_chi_eta=59.254199}
		\flushright
		\includegraphics[width=\textwidth]{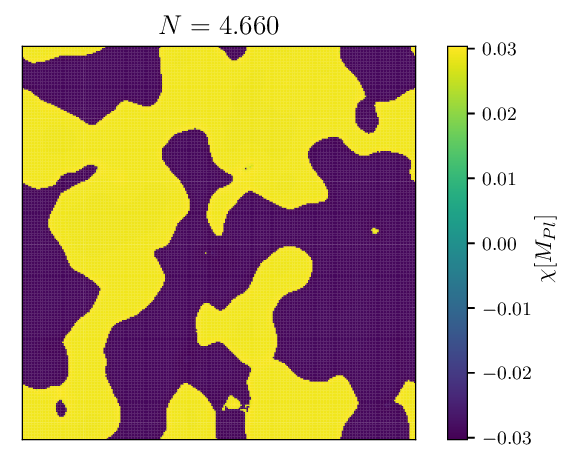}
	\end{subfigure} 
    \caption{Snapshots of the spatial distribution
	  of the spectator field $\chi$ before (upper panels), during (middle panels) and after (lower panels) geometrical destabilization
	  on a section of the lattice. The plots are order in increasing time from left to right and from top to bottom. All results are given for model 1A listed in Table~\ref{tab:pars}.}
    \label{fig:dump_chi}
\end{figure*}

\begin{figure*}[!ht]
	\begin{subfigure}[t]{0.5\textwidth}
		\label{fig:parameters_alpha=1e-3_n=2_plot_barotropic}
		\flushleft
		\includegraphics[width=215pt]{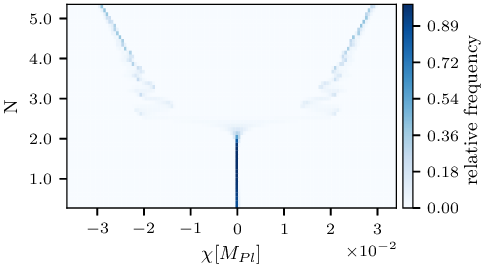}
	\end{subfigure}\hfill %
	\begin{subfigure}[t]{0.5\textwidth}
		\label{fig:parameters_alpha=1e-3_n=3_plot_barotropic}
		\flushright
		\includegraphics[width=215pt]{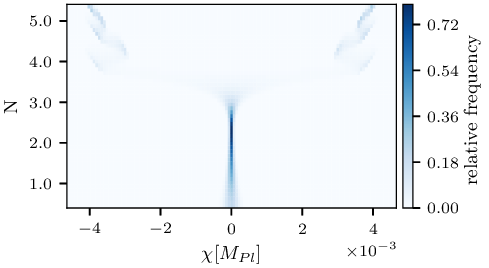}
	\end{subfigure} \\
	\begin{subfigure}[t]{0.5\textwidth}
		\label{fig:parameters_alpha=1e-4_n=2_plot_barotropic}
		\flushleft
		\includegraphics[width=215pt]{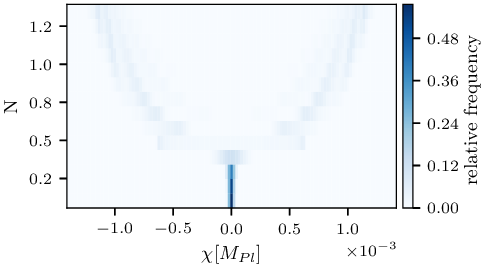}
	\end{subfigure}\hfill %
	\begin{subfigure}[t]{0.5\textwidth}
		\label{fig:parameters_alpha=1e-4_n=3_plot_barotropic}
		\flushright
		\includegraphics[width=215pt]{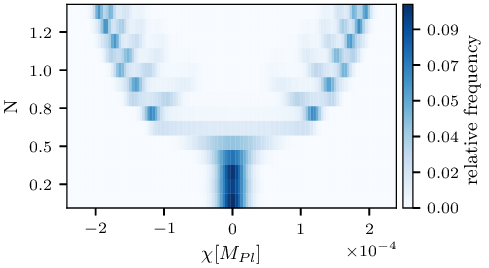}
	\end{subfigure} %
	  \caption{Time evolution of the distribution of the amplitude of $\chi$ in nodes of the lattice for the models listed in Table~\ref{tab:pars}. Field values from the range displayed in the plots are binned and the shade of the bin corresponds to
	  the proportion of nodes at which the field value correspond
	  to a given bin.}
    \label{fig:dist_chi}
\end{figure*}

We have performed numerical lattice simulations based on numerical algorithm described in Appendix \ref{app:numerical_scheme}. In our computations we used lattices of the size of $512^3$ lattice points, with the exception for plots in figures \ref{fig:dump_chi} and \ref{fig:dist_chi} for which results from simulations on lattices with $256^3$ points were used in order to reduce storage space and time of processing. For each set of parameters listed in Table \ref{tab:pars} average of 5 independent simulations was used in the analysis. Moreover, many more trail simulations were performed in order to determine optimal value of the lattice spacing $h$, i.e. comoving distance between neighbouring lattice points, given in Table \ref{tab:pars} as cutoff scale $k_{\rm cutoff} = \sqrt{3} \pi / h$. Comparing results of simulations with values of $k_{\rm cutoff}$ that differ by the factor of $2$ we were able to study the effects of spatial discretization of equations of motion and the role of IR and UV cutoffs that are inevitably associated with lattice simulations. 

As usually we chose the Bunch-Davies initial conditions for perturbations which are appropriate for quantum fields in time-dependent, de Sitter background. This procedure is well-known for fields with trivial field-space metric and was generalized for non-trivial cases and prescription for initial conditions of the perturbations can be found e.g.\ in \cite{Lalak:2007vi}. Exactly the same method was used for initialization of simulations of geometrical destabilization during preheating in $\alpha$-attractor models of inflation in \cite{Krajewski:2018moi, Wieczorek:2019umc, Krajewski:2022ezo} where it was described in detail.

\begin{figure}[h]
    \centering
    \includegraphics[width=0.45\textwidth]{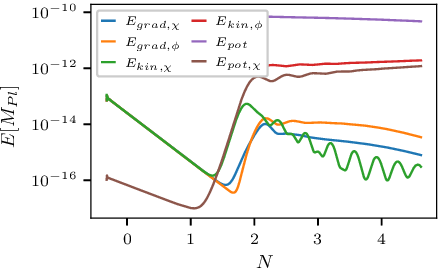}
    \includegraphics[width=0.45\textwidth]{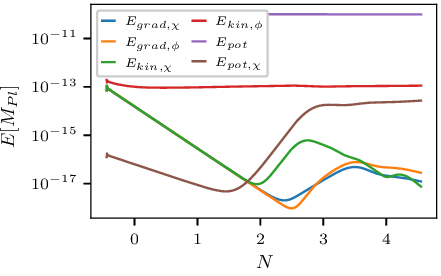} \\
    \includegraphics[width=0.45\textwidth]{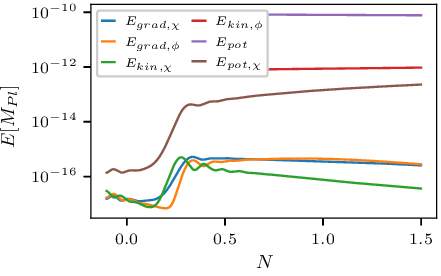}
    \includegraphics[width=0.45\textwidth]{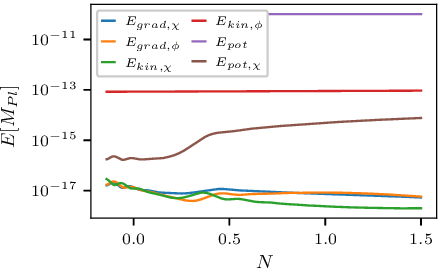}
    \caption{Evolution of various components of the energy density for the models listed in Table~\ref{tab:pars}. There are six components, corresponding to the kinetic, gradient and potential
    energy of each of fields $\phi$ and $\chi$.}
    \label{fig:evo_energy}
\end{figure}

Perhaps the most instructive way of presenting our results consists in
showing the time evolution of various components of the total energy density, as illustrated in the Figure~\ref{fig:evo_energy}. We show kinetic and  gradient energy density of each of fields $\phi$ and $\chi$, as well as the total potential energy density and the potential energy density for $\chi$. The separation of the kinetic and gradient energy density into components corresponding to each field is possible because the metric of
the field space is diagonal. The separation of the potential energy density into components corresponding to each field could be done, because the potential is a sum of two independent contributions, one from each field.
The onset of geometrical destabilization, corresponding to the moment
at which the effective mass (\ref{eq:ms2_2}) changes sign to negative
is marked as $N=0$. 

We observe that quite soon after geometrical destabilization begins, 
the kinetic and potential energy density of $\chi$ starts to grow,
and that this growth is soon followed, but not matched, 
by gradient energy densities of the fields. This shows that
spatial inhomogeneities have a subdominant role in the evolution
of the Universe. The growth of tachyonic instabilities is very short
and it is quickly terminated by backreaction effects.

\begin{figure}[h]
    \centering
    \includegraphics[width=0.45\textwidth]{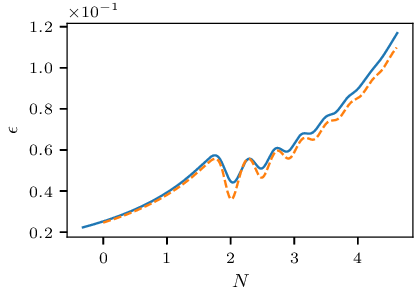}
    \includegraphics[width=0.45\textwidth]{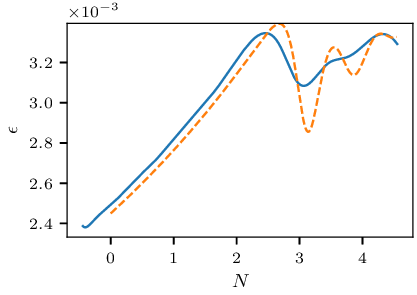} \\
    \includegraphics[width=0.45\textwidth]{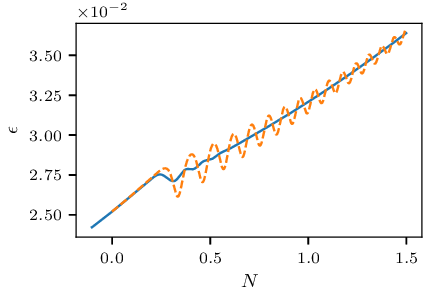}
    \includegraphics[width=0.45\textwidth]{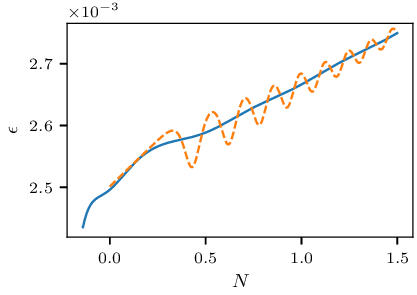}
    \caption{Evolution of the slow roll parameter $\epsilon_1$ for the models listed in Table~\ref{tab:pars}. Solid blue lines correspond to
    the results of lattice simulations. Dashed orange lines correspond to the approximate analysis presented in Ref.~\cite{Grocholski:2019mot}.}
    \label{fig:evo_epsilon}
\end{figure}

In Ref.~\cite{Grocholski:2019mot}, it was hypothesized that
the backreaction was mainly caused by the growth of amplitudes
of perturbations of the field~$\chi$, which led to a 
geometrical suppression of the curvature term in the effective
mass (\ref{eq:ms2_2}). This conclusion was obtained by following
equations of motion for homogeneous fields,
with initial conditions with the displacement of the field strength~$\chi$
from zero to a~value inferred from stochastic inflation considerations.

While this method captures the characteristic features
of the field dynamics -- fast growth followed by oscillations
around a slow-roll trajectory corresponding to partial equilibrium
between geometrical effects and the pull of the potential --
our numerical studies do not confirm this picture on the finer,
quantitative level. In Figure~\ref{fig:evo_epsilon} we show
the evolution of the slow-roll parameter~$\epsilon_1$, superimposing the results of our lattice simulations (solid blue lines) with the approximate method of \cite{Grocholski:2019mot}.
It is clearly visible that the backreaction effects become important
earlier than the approximate analysis suggests and 
that the effects of geometrical destabilization,
already known to be tamed down by backreaction, are even
weaker than anticipated by the approximate analysis.

\begin{figure}[h]
    \centering
    \includegraphics[width=0.45\textwidth]{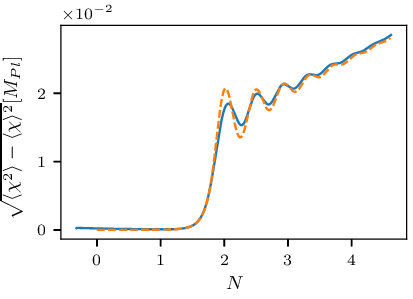}
    \includegraphics[width=0.45\textwidth]{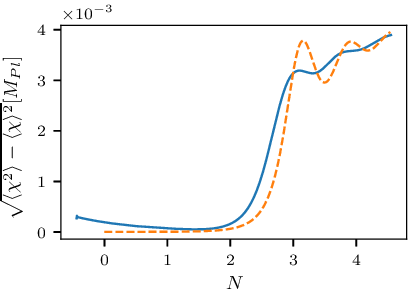} \\
    \includegraphics[width=0.45\textwidth]{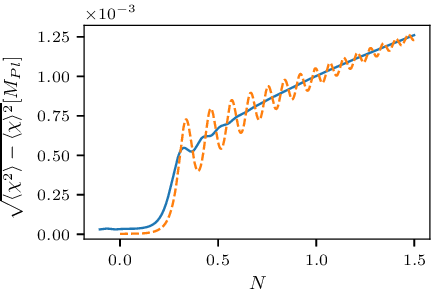}
    \includegraphics[width=0.45\textwidth]{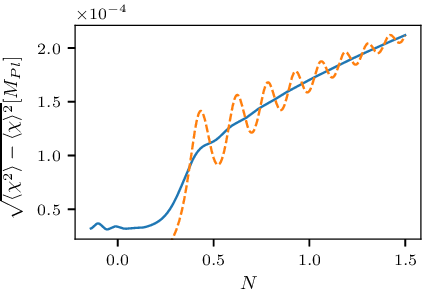}
    \caption{Evolution of the standard deviation of $\chi$ and for the models listed in Table~\ref{tab:pars}. Dashed orange lines correspond to the results for the average value of $\chi$ obtained from the approximate analysis presented in Ref.~\cite{Grocholski:2019mot}.}
    \label{fig:evo_chi}
\end{figure}

Similar conclusions can be drawn from the evolution of $\chi$. In Figure~\ref{fig:evo_chi}, we present standard deviation $\sqrt{\langle(\chi - \langle \chi \rangle)^2\rangle}$ of this field strength\footnote{Strictly speaking, $\langle \chi \rangle$ should be equal to zero when averaged over the lattice volume and many realizations of simulations due to $Z_2$ symmetry of the considered model. For a single simulation, finite-volume effects lead to $0< \langle \chi \rangle^2\ll \langle \chi^2 \rangle$.}, and compare it with the results obtained with the approximate method of \cite{Grocholski:2019mot}. 
Comparing the standard deviation with the average field value,
 we note that in all cases the former is much larger than the latter, which may suggest the formation of patches in space, with values of
 $\chi$ differing in sign. Additionally, the approximate method underestimates the fluctuations growth at the early stages of
 the geometrical destabilization and overestimates the range of oscillations of those fluctuations. 

\begin{figure*}[!ht]
\centering
	\begin{subfigure}[t]{0.5\textwidth}
		\label{fig:plot_hist_chi_eta=0.000891}
		\flushleft
		\includegraphics[width=\textwidth]{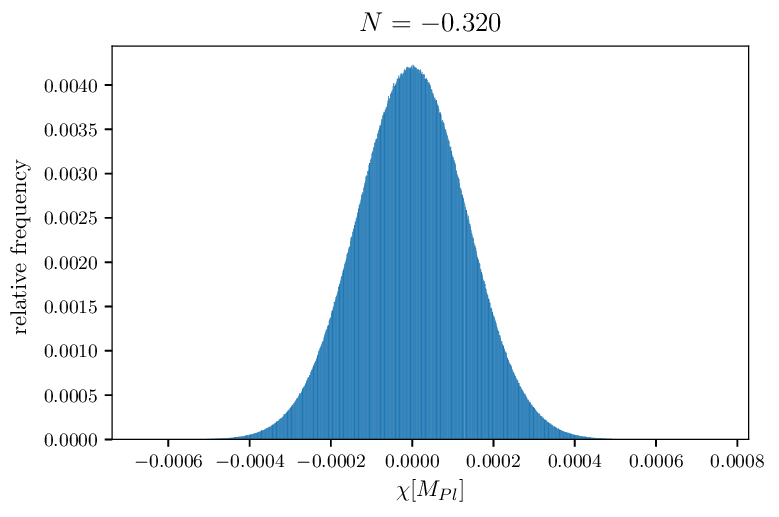}
	\end{subfigure}\hfill %
	\begin{subfigure}[t]{0.5\textwidth}
		\label{fig:plot_hist_chi_eta=47.156771}
		\flushright
		\includegraphics[width=\textwidth]{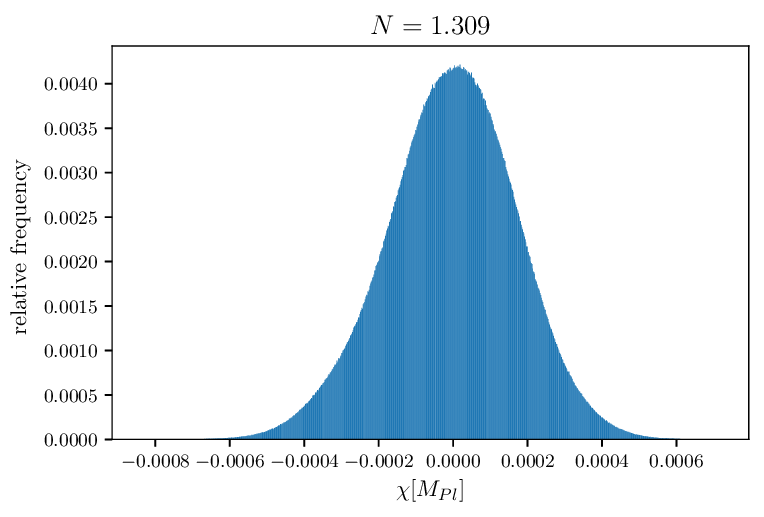}
	\end{subfigure}
    \\
	\begin{subfigure}[t]{0.5\textwidth}
		\label{fig:plot_hist_chi_eta=51.758021}
		\flushleft
		\includegraphics[width=\textwidth]{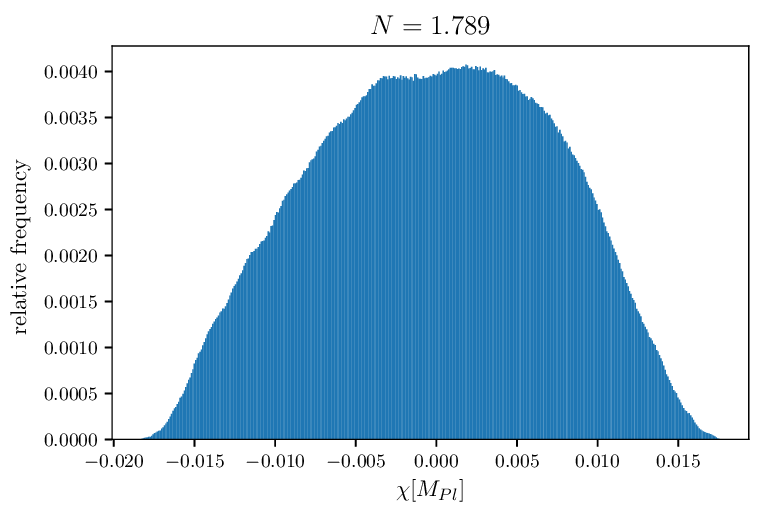}
	\end{subfigure}\hfill %
	\begin{subfigure}[t]{0.5\textwidth}
		\label{fig:plot_hist_chi_eta=52.453381}
		\flushright
		\includegraphics[width=\textwidth]{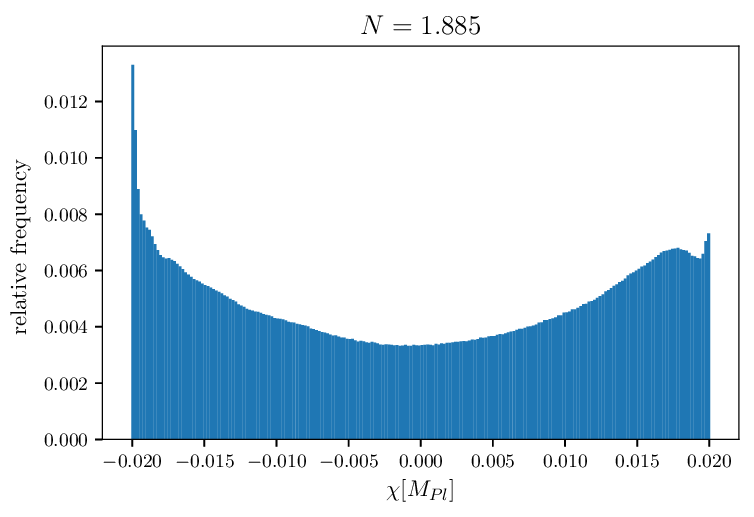}
	\end{subfigure}
    \\
    \begin{subfigure}[t]{0.5\textwidth}
		\label{fig:plot_hist_chi_eta=55.114261}
		\flushleft
		\includegraphics[width=\textwidth]{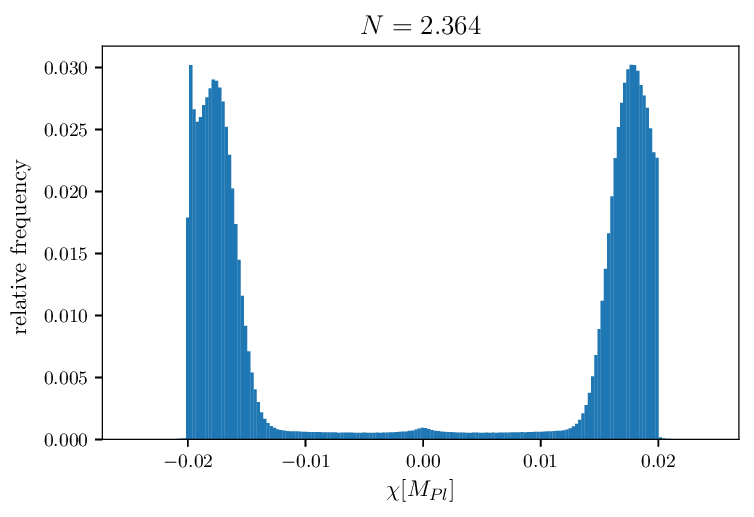}
	\end{subfigure}\hfill %
	\begin{subfigure}[t]{0.5\textwidth}
		\label{fig:plot_hist_chi_eta=59.254199}
		\flushright
		\includegraphics[width=\textwidth]{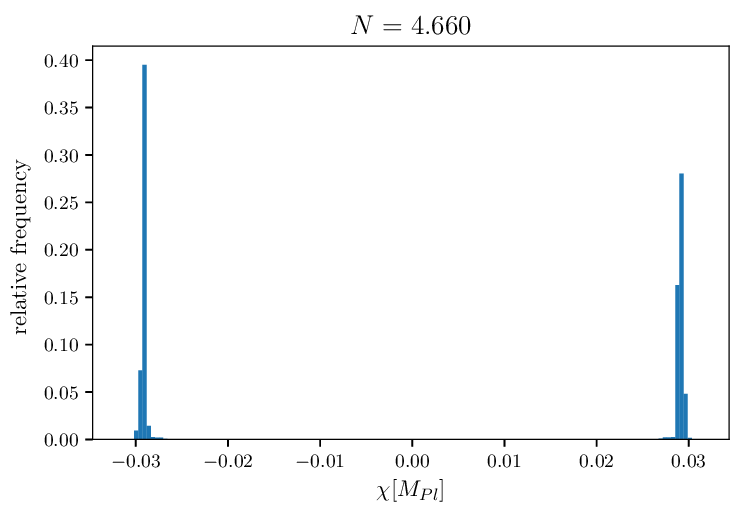}
	\end{subfigure} %
	  \caption{Histograms of relative
	  frequency of different values of the spectator field $\chi$ before (upper panels), during (middle panels) and after (lower panels) geometrical destabilization on a section of the lattice. The plots are order in increasing time from left to right and from top to bottom. All results are given for model 1A listed in Table~\ref{tab:pars}.}
    \label{fig:hist_chi}
\end{figure*}

In order to corroborate the formation of patches in the Universe,
corresponding to approximately uniform values of the displaced $\chi$ field, in Figure~\ref{fig:dump_chi} we show the distribution
of the values of the spectator field $\chi$ and in Figure~\ref{fig:hist_chi} the histograms of relative probability density of finding $\chi$ close to
a particular value for six different instances,
corresponding to times before, during and after
geometrical destabilization. We can see that an initial
random distribution is transformed to one that exhibits
discernible patches
corresponding to different signs but the same overall amplitude of
$\chi$, and separated by domain walls.
We can also determine that the size of these patches 
roughly
corresponds to the
Hubble radius at the onset of geometrical 
destabilization.\footnote{This effect was first reported based on a much smaller-scale simulations in \cite{Wieczorek:2019umc}.} 
Note that the apparent small breaking of $Z_2$ symmetry of
the distributions in Figure~~\ref{fig:hist_chi} stems from
the fact that we report a single realization of a simulation
with stochastic initial conditions, so it represents
the inevitable cosmic variance.

In order to analyze further the formation of patches in the Universe,
corresponding to approximately uniform values of the displaced $\chi$ field,
we studied how the distribution of the amplitudes of $\chi$ in 
nodes of the lattice evolves in time. The results are presented 
in Figure~\ref{fig:dist_chi}. We can see that before the onset
of geometrical destabilization the most probable field value
is peaked at zero; later, the evolution of the field bifurcates and
the distribution becomes bimodal, with two sharp peaks corresponding
to an almost constant field value within each patch, 
and that neighboring patches differ in the sign of $\chi$,
while the overall amplitude of the field is the same.

\section{Discussion and conclusions}
\label{sec:discussion}

When geometrical destabilization of inflation
was originally advocated \cite{RT},
it was hypothesized that there could be two possible outcomes of this
instability. Either the curvature perturbations could grow
uncontrollably, effectively terminating inflation, or their
evolution could be much less dramatic, with a shift of the spectator
field to a new value determined by a balance between the
divergence of the geodesic lines (producing an apparent force
akin to apparent forces in a non-inertial reference frame) and 
the force resulting from the field potential. 

The first option
was adopted in \cite{GD1}, which provided a Bayesian
analysis of the parameter space of the inflationary models
under the assumption of a sudden termination of inflation. 

A radically different view was put forth in \cite{GD2}, which
assumed that the spectator field is merely `side-tracked'
to a new classical state and that after geometrical destabilization
inflation proceeds along a new classical trajectory. 
Inflation along this trajectory can give rise to a number of interesting
phenomena, e.g. a transient instability 
of the curvature perturbations related to a negative
effective sound speed \cite{Cremonini:2010ua} (see also
\cite{Brown:2017osf,Mizuno:2017idt}) and a peculiar shape and
amplitude of generated non-Gaussianities \cite{GD2}.
Similarly, it was suggested in
\cite{Cicoli:2018ccr}
that geometrical destabilization of inflation
must be terminated, because the would-be spectator
mass evaluated on the attractor solution is always positive. 

The arguments recapitulated above are based on the analysis 
of a homogeneous inflationary background, so it is unclear
whether they can be applicable once geometrical destabilization
is in place. A step further was made in \cite{Grocholski:2019mot},
showing a negative feedback loop intrinsically built into any
model of geometrical destabilization. What remained elusive
at that point was a fully numerical corroboration of
that observation.

Our work fills what we believe was 
the last gap in the discussion of geometrical
destabilization of inflation.
Here we have performed state-of-the-art lattice
simulations to show that that the instability is quickly shut
off by backreaction effects and that they are even 
stronger than previously anticipated.
Our calculations have conclusively shown that geometrical
destabilization is a short-lasting phenomenon and that its
main consequence consists in the fact that the classical
fields leave the configuration which is no longer stable and
move, within a causally connected patch, to a new, stable one.

Our numerical simulations show formation of domain walls. 
It is a~natural question how these structure evolve during reheating
and subsequent radiation domination phase. The emergence of domain
walls in our model is related to the $Z_2$-symmetric kinetic coupling 
between the spectator field and the velocity of the inflaton field.
It is known that in models with a large negative curvature of the field
space, inflaton oscillations around the minimum of the potential can
lead to fragmentation of both the inflaton and the spectator
field \cite{Krajewski:2018moi, Krajewski:2022ezo}. 
As a result, one can envision either an additional bout
of production of such structures during reheating or,
conversely, their destruction.
A robust verification of any such scenario will require dedicated multi-scale numerical study and is beyond scope of this manuscript. 
Assuming that the domain walls related to the spectator field and produced during the geometrical phase of inflation survived reheating, we can
predict their observational consequences with less uncertainty. If potential of the model has only one minimum that preserves the symmetry, 
the domains will decay via the misalignment mechanism 
toward minimum of the potential, 
producing quanta of the spectator field. 
If the potential has symmetry breaking minima, the evolution of domain walls can be more complicated. In such a~case, structures produced during inflation will probably form a~network of domain walls, superficially
resembling systems that have been discussed widely in the literature \cite{Lalak:2007rs, Coulson:1995uq, Press:1989yh}-- with a crucial difference that sizes of domains are very large comparing to their wall widths -- a result of stretching during inflation. One may expect that the network will enter the scaling regime \cite{Leite:2011sc, Avelino:2005pe, Oliveira:2004he, Garagounis:2002kt} soon after the horizon will reach the size of the order of the domains. 
Such network will cause a number of 
well-known problems \cite{Martins:2016ois, Lazanu:2015fua, Conversi:2004pi, Friedland:2002qs, Fabris:2000qz, Zeldovich:1974uw}, unless it is destabilized by one of two mechanisms: 
a tilt in the potential that softly breaks the $Z_2$ symmetry \cite{Correia:2018tty, Correia:2014kqa, Avelino:2008qy}
or a probability bias \cite{Krajewski:2021jje, Casini:2001ai, Larsson:1996sp, Lalak:1996db, Hindmarsh:1996xv, Coulson:1995nv, Lalak:1994qt, Gelmini:1988sf}. 
With large enough symmetry breaking, the former mechanism can be effective enough
to force the decay of the network before it enters scaling regime. 
The determination if the bias is a~viable mechanism of destabilizing the network will require separate dedicated study, so we leave it for future research. 
Decaying domain walls will produce both quanta of the spectator field and gravitational waves \cite{Hiramatsu:2013qaa, Kawasaki:2011vv}.
Therefore, in our view, the emergence of the domain walls 
seems to be by far the most interesting consequence of
geometrical destabilization, as it provokes a number of 
difficult questions that can seed future research.

\subsubsection*{Acknowledgements}

We thank Micha\l\ Wieczorek for collaboration at
an early stage of this project.
T.K.~is supported by grant 2019/32/C/ST2/00248 from 
the National Science Centre (Poland).
K.T.~is partially supported by grant 2018/30/Q/ST9/00795 
from the National Science Centre (NCN).
This research was supported in part by PL--Grid Infrastructure. 
T.K.\ acknowledges hospitality of Rudolf Peierls Centre for Theoretical Physics at Oxford University, where parts of this work has been done. 
K.T.\ thanks International Physicists' Tournament at Universidad Industrial de Santander in Bucaramanga
(Colombia) for fruitful discussions and 
stimulating atmosphere during the last stages of the
completion of this work.

\begin{appendix}

\section{Numerical discretization scheme for lattice simulations\label{app:numerical_scheme}}

The method presented here is a~slight modification of the method that we used for studies of dynamics of preheating in $\alpha$-attractor T-models of inflation in \cite{Krajewski:2018moi,Krajewski:2022ezo}. Only non-canonical kinetic term coupling $e^{2b(\chi)}$ and the potential $V(\phi, \chi)$ had to be modified. For the model under consideration these functions are:
\begin{align}
    e^{2b(\chi)} &= 1 + \frac{2\chi^2}{M_{\text{np}}^2}, & V(\phi, \chi) &= V(\phi) + \frac{1}{2}m_h^2 \chi^2.
\end{align}
In the current section we will present the algorithm in its general formulation.

\subsection{Lattice formulation of the model}

We consider the general model described by the following action
\begin{equation}\label{action2}
	S=\int d^4x\sqrt{-g}\bigg[\frac{{M_{Pl}}^2}{2}\mathcal{R}-\frac{1}{2}e^{2b(\chi)}(\partial_\mu\phi)(\partial^\mu\phi)-\frac{1}{2}(\partial_\mu\chi)(\partial^\mu\chi)-V(\phi,\chi)\bigg].
\end{equation}

We assume here that the spacetime is spatially homogeneous, isotropic and flat, thus we approximate the metric tensor field by Friedman-Lema\^itre-Robertson-Walker metric expressed in terms of the conformal time $\tau$:
\begin{equation}
	ds^2=a^2(-d\tau^2+d\mathbf{x}^2),
\end{equation}
which implies
\begin{equation}
	\sqrt{-g}=a^4\qquad \textrm{and}\qquad \mathcal{R}=6\frac{a''}{a^3},
\end{equation}
where the prime denotes the derivative with respect to $\tau$.
In our simulations we neglected the backreaction from metric fluctuations on the evolution of both the inflaton and the spectator, still keeping tack of the evolution of the scale factor $a$ of the background metric.

Our aim is to obtain symplectic integrator for model under consideration. We use the method of lines, i.e. first we discretized the action in space and then we constructed the symplectic integrator for the theory defined on the lattice. We simulate some patch of the Universe with finite comoving volume $U$, thus discrete theory has only finite number of degrees of freedom. Under the assumption of homogeneity of the Universe it is natural to use periodic boundary conditions. 

We use values of the field strengths of $\chi$ and $\phi$ at the nodes of cubic regular lattice as degrees of freedom for discrete theory. The spacial gradients of fields can be approximated using finite difference method. In our implementation we used first order forward finite difference scheme. This choice is equivalent to linear interpolation of  field strength values at cells of the lattice.

After discretization in space the action \eqref{action2} can be written as:
\begin{eqnarray}\nonumber
	S & = & \int L d\tau =\\
	{} & = & \int \bigg[-3a'^2 U {M_{Pl}}^2 +\sum_{\vec{x}}\frac{a^2}{2}\frac{U}{V_L}\Bigg(e^{2b(\chi_{\vec{x}})}\bigg((\phi'_{\vec{x}})^2-G_{\vec{x}}(\phi)\bigg)+\\ \nonumber
	{} & + & \bigg((\chi'_{\vec{x}})^2-G_{\vec{x}}(\chi)\bigg)-a^2V(\phi_{\vec{x}},\chi_{\vec{x}})\Bigg)\bigg]d\tau,
\end{eqnarray}
where $V_L$ is the number of cells in the lattice, thus $\frac{U}{V_L}$ is equal to the volume of each cell, $\vec{x} = (x_1, x_2, x_3)^T \colon x_1, x_2, x_3 \in \mathbb{N}$ is a~multi-index that numerates lattice points in three dimensions and
\begin{equation}
	G_{\vec{x}}(Y)=\frac{1}{2 \delta^2}  \sum_{i = 1}^{3}(Y_{\vec{x} + \vec{e_i}}-Y_{\vec{x}})^2
\end{equation}
is the square of discretization of the spatial gradient with $\vec{e_1} = (1, 0, 0)^T$, $\vec{e_2} = (0, 1, 0)^T$ and $\vec{e_3} = (0, 0, 1)^T$ and $\delta$ is comoving length of the edge of a~lattice cell.

After dropping common factor $\frac{U}{V_L}$ and the Legendre transformation we obtain the following Hamiltonian:
\begin{equation}
	H=-\frac{p_a^2}{12V_L {M_{Pl}}^2}+\sum_{\vec{x}}a^4\bigg(\frac{\pi^2_{\phi,\vec{x}}}{2a^6e^{2b(\chi_{\vec{x}})}}+
	\frac{\pi^2_{\chi,\vec{x}}}{2a^6}+e^{2b(\chi_{\vec{x}})}\frac{G_{\vec{x}}(\phi)}{2\delta^2 a^2}+\frac{G_{\vec{x}}(\chi)}{2 \delta^2 a^2}+V(\phi_{\vec{x}},\chi_{\vec{x}})\bigg),
\end{equation}
where canonical momenta are defined by formulae:
\begin{equation}
	p_a\equiv\frac{\partial L}{\partial a'}=-6a'V_L {M_{Pl}}^2,\quad \pi_{\phi,\vec{x}}\equiv \frac{\partial L}{\partial \phi'}=a^2\textrm{e}^{2b(\chi_{\vec{x}})}\phi'_{\vec{x}}\quad\textrm{and}\quad \pi_{\chi,\vec{x}}\equiv \frac{\partial L}{\partial \chi'}=a^2\chi'_{\vec{x}}.
\end{equation}

From the point of view of implementation of numerical algorithm it is convenient to use 
\begin{equation}
\wp_a := \frac{p_a}{V_L} = -6a'{M_{Pl}}^2.
\end{equation}
instead of canonical momentum $p_a$.

\subsection{Time integration scheme}
Construction of the time integrator that we used in our simulations is based on the technique of operator splitting. The Hamiltonian $H$ can be divided into four parts
\begin{equation}
	H=H_1+H_2+H_3+H_4
\end{equation}
in such the way that Hamilton's equations for each part can be integrated explicitly. We used following splitting:
\begin{equation}
	H_1\equiv -\frac{p_a^2}{12V_L {M_{Pl}}^2},
\end{equation}
\begin{equation}
	H_2\equiv \sum_{\vec{x}}a^4\bigg(\frac{\pi^2_{\phi,\vec{x}}}{2a^6e^{2b(\chi_{\vec{x}})}}\bigg),
\end{equation}
\begin{equation}
	H_3\equiv\sum_{\vec{x}}a^4\bigg(\frac{\pi^2_{\chi,\vec{x}}}{2a^6}\bigg)
\end{equation}
and
\begin{equation}
	H_4\equiv \sum_{\vec{x}}a^4\bigg(e^{2b(\chi_{\vec{x}})}\frac{G_{\vec{x}}(\phi)}{2 \delta^2a^2}+\frac{G_{\vec{x}}(\chi)}{2 \delta^2a^2}+V(\phi_{\vec{x}},\chi_{\vec{x}})\bigg).
\end{equation}
The corresponding flows for time step $\delta \tau$ are as follows:
\begin{equation}
	\Phi_1(h):\bigg(a,p_a,\phi_{\vec{x}},\pi_{\phi,\vec{x}},\chi_{\vec{x}},\pi_{\chi,\vec{x}}\bigg)\rightarrow \bigg(a+\frac{\partial H_1}{\partial p_a}\delta \tau ,p_a,\phi_{\vec{x}},\pi_{\phi,\vec{x}},\chi_{\vec{x}},\pi_{\chi,\vec{x}}\bigg),
\end{equation}
\begin{equation}
	\Phi_2(h):\bigg(a,p_a,\phi_{\vec{x}},\pi_{\phi,\vec{x}},\chi_{\vec{x}},\pi_{\chi,\vec{x}}\bigg)\rightarrow \bigg(a,p_a-\frac{\partial H_2}{\partial a}\delta \tau,\phi_{\vec{x}}+\frac{\partial H_2}{\partial \pi_{\phi,\vec{x}}}\delta \tau,\pi_{\phi,\vec{x}},\chi_{\vec{x}},\pi_{\chi,\vec{x}}-\frac{\partial H_2}{\partial \chi_{\vec{x}}}\delta \tau\bigg),
\end{equation}
\begin{equation}
	\Phi_3(h):\bigg(a,p_a,\phi_{\vec{x}},\pi_{\phi,\vec{x}},\chi_{\vec{x}},\pi_{\chi,\vec{x}}\bigg)\rightarrow \bigg(a,p_a-\frac{\partial H_3}{\partial a}\delta \tau,\phi_{\vec{x}},\pi_{\phi,\vec{x}},\chi_{\vec{x}}+\frac{\partial H_3}{\partial \pi_{\chi,\vec{x}}}\delta \tau,\pi_{\chi,\vec{x}}\bigg)
\end{equation}
and
\begin{equation}
	\Phi_4(h):\bigg(a,p_a,\phi_{\vec{x}},\pi_{\phi,\vec{x}},\chi_{\vec{x}},\pi_{\chi,\vec{x}}\bigg)\rightarrow \bigg(a,p_a-\frac{\partial H_4}{\partial a}\delta \tau,\phi_{\vec{x}},\pi_{\phi,\vec{x}}-\frac{\partial H_4}{\partial \phi_{\vec{x}}}\delta \tau,\chi_{\vec{x}},\pi_{\chi,\vec{x}}-\frac{\partial H_4}{\partial \chi_{\vec{x}}}\delta \tau\bigg).
\end{equation}

Using Strang splitting one can easily construct second order symplectic method which we use in our numerical simulations:
\begin{equation}
	\Phi(h) =\Phi_1(h/2)\circ\Phi_2(h/2)\circ\Phi_3(h/2)\circ\Phi_4(h)\circ\Phi_3(h/2)\circ\Phi_2(h/2)\circ\Phi_1(h/2).
\end{equation}
It can be written as the following set of explicit upgrades:
\begin{equation}
a_{n+1/2}=a_n+\frac{\delta \tau}{2} {V_L}^{-1} \frac{\partial\tilde{H}_1}{\partial \wp_a}(\wp_{a,n})
\end{equation}
\begin{equation}
\tilde{\wp}_{a,n+1/2}=\wp_{a,n}-\frac{\delta \tau}{2} {V_L}^{-1} \frac{\partial H_2}{\partial a}(a_{n+1/2},\pi_{\phi,\vec{x},n},\chi_{\vec{x},n})
\end{equation}
\begin{equation}
\tilde{\pi}_{\chi,\vec{x},n+1/2}=\pi_{\chi,\vec{x},n}-\frac{\delta \tau}{2}\frac{\partial H_2}{\partial \chi_{\vec{x}}}(a_{n+1/2},\pi_{\phi,\vec{x},n},\chi_{\vec{x},n})
\end{equation}
\begin{equation}
\phi_{\vec{x},n+1/2}=\phi_{\vec{x},n}+\frac{\delta \tau}{2}\frac{\partial H_2}{\partial \pi_{\phi,\vec{x}}}(a_{n+1/2},\pi_{\phi,\vec{x},n},\chi_{\vec{x},n})
\end{equation}
\begin{equation}
\tilde{\tilde{\wp}}_{a,n+1/2}=\tilde{\wp}_{a,n+1/2}-\frac{\delta \tau}{2} {V_L}^{-1} \frac{\partial H_3}{\partial a}(a_{n+1/2},\tilde{\pi}_{\chi,\vec{x},n+1/2})
\end{equation}
\begin{equation}
\chi_{\vec{x},n+1/2}=\chi_{\vec{x},n}+\frac{\delta \tau}{2}\frac{\partial H_3}{\partial \pi_{\chi,\vec{x}}}(a_{n+1/2},\tilde{\pi}_{\phi,\vec{x},n+1/2})
\end{equation}
\begin{equation}
\tilde{\tilde{\wp}}_{a,n+1}=\tilde{\tilde{\wp}}_{a,n+1/2}-h {V_L}^{-1} \frac{\partial H_4}{\partial a}(a_{n+1/2},\phi_{\vec{x},n+1/2},\chi_{\vec{x},n+1/2})
\end{equation}
\begin{equation}
\pi_{\phi,\vec{x},n+1}=\pi_{\phi,\vec{x},n}-\delta \tau\frac{\partial H_4}{\partial \phi{\vec{x}}}(a_{n+1/2},\phi_{\vec{x},n+1/2},\chi_{\vec{x},n+1/2})
\end{equation}
\begin{equation}
\tilde{\pi}_{\chi,\vec{x},n+1}=\tilde{\pi}_{\chi,\vec{x},n+1/2}-\delta \tau\frac{\partial H_4}{\partial \chi}(a_{n+1/2},\phi_{\vec{x},n+1/2},\chi_{\vec{x},n+1/2})
\end{equation}
\begin{equation}
\chi_{\vec{x},n+1}=\chi_{\vec{x},n+1/2}+\frac{\delta \tau}{2}\frac{\partial H_3}{\partial \pi_{\chi,\vec{x}}}(a_{n+1/2},\tilde{\pi}_{\phi,\vec{x},n+1})
\end{equation}
\begin{equation}
\tilde{\wp}_{a,n+1}=\tilde{\tilde{\wp}}_{a,n+1}-\frac{\delta \tau}{2} {V_L}^{-1} \frac{\partial H_3}{\partial a}(a_{n+1/2},\tilde{\pi}_{\chi,\vec{x},n+1})
\end{equation}
\begin{equation}
\phi_{\vec{x},n+1}=\phi_{\vec{x},n+1/2}+\frac{h}{2}\frac{\partial H_2}{\partial \pi_{\phi,\vec{x}}}(a_{n+1/2},\pi_{\phi,\vec{x},n+1},\chi_{\vec{x},n+1})
\end{equation}
\begin{equation}
\pi_{\chi,\vec{x},n+1}=\tilde{\pi}_{\chi,\vec{x},n+1}-\frac{\delta \tau}{2}\frac{\partial H_2}{\partial \chi_{\vec{x}}}(a_{n+1/2},\pi_{\phi,\vec{x},n+1},\chi_{\vec{x},n+1})
\end{equation}
\begin{equation}
\wp_{a,n+1}=\tilde{\wp}_{a,n+1}-\frac{\delta \tau}{2} {V_L}^{-1} \frac{\partial H_2}{\partial a}(a_{n+1/2},\pi_{\phi,\vec{x},n+1},\chi_{\vec{x},n+1})
\end{equation}
\begin{equation}
a_{n+1}=a_{n+1/2}+\frac{\delta \tau}{2} {V_L}^{-1} \frac{\partial\tilde{H}_1}{\partial \wp_a}(\wp_{a,n+1})
\end{equation}
where we have defined:
\begin{equation}
\tilde{H}_1 (\wp_a) := H_1 (V_L \wp_a) = -\frac{V_L \wp_a^2}{12 {M_{Pl}}^2}.
\end{equation}

\end{appendix}

\end{document}